\documentclass[aps,prb,twocolumn,groupedaddress,showpacs,superscriptaddress,amssymb,amsmath]{revtex4-1}
\usepackage{graphicx}
\usepackage{amsmath}
\usepackage{amssymb}
\usepackage{amsfonts}
\usepackage{color}
\usepackage{bm}        
\usepackage{comment}
\usepackage{placeins}
\usepackage[mathscr]{eucal}
\usepackage[colorlinks, linkcolor=red,citecolor=blue,urlcolor=blue]{hyperref}
\usepackage[normalem]{ulem}
\usepackage[all]{hypcap}
\usepackage{multirow}
\usepackage[dvipsnames,usenames,table]{xcolor}
\usepackage{dcolumn}
\usepackage{hyperref}
\usepackage{bm}
\usepackage{epsf}
\usepackage{subfigure}
\usepackage{amsfonts}
\usepackage{epstopdf}%
\newcommand{\be}{\begin{equation}}
\newcommand{\ee}{\end{equation}}
\newcommand{\bea}{\begin{eqnarray}}
\newcommand{\eea}{\end{eqnarray}}
\begin{document}
\title{Thermodynamic uncertainty relation in atomic-scale quantum conductors}

\author{Hava Meira Friedman}
\affiliation{Chemical Physics Theory Group, Department of Chemistry, 
University of Toronto, 80 Saint George St., Toronto, Ontario, Canada M5S 3H6}
\author{Bijay K. Agarwalla}
\affiliation{Department of Physics, Dr. Homi Bhabha Road, Indian Institute of Science Education and Research, Pune 411008, India}
\author{Ofir Shein-Lumbroso} 
\affiliation{Department of Chemical and Biological Physics, Weizmann Institute of Science, Rehovot, Israel}
\author{Oren Tal}
\affiliation{Department of Chemical and Biological Physics, Weizmann Institute of Science, Rehovot, Israel}
\author{Dvira Segal}
\email{dvira.segal@utoronto.ca}
\affiliation{Chemical Physics Theory Group, Department of Chemistry,
University of Toronto, 80 Saint George St., Toronto, Ontario, Canada M5S 3H6}
\affiliation{Department of Physics, University of Toronto, Toronto, Ontario, Canada M5S 1A7}



\date{\today}

\begin{abstract}
The thermodynamic uncertainty relation (TUR) is expected to hold
in nanoscale electronic conductors, when the electron transport process is quantum coherent and the transmission probability is 
constant (energy and voltage independent). 
We present measurements of the electron current and its noise in gold atomic-scale junctions and confirm the validity of 
the TUR for electron transport in realistic quantum coherent conductors. 
Furthermore, we show that it is beneficial to present the current and its noise as a 
TUR ratio in order to identify deviations from noninteracting-electron coherent dynamics.
\end{abstract}

\maketitle

\section{Introduction}
\label{Sintro}


The thermodynamic uncertainty relation (TUR), a cost-precision trade-off relationship,
has been of great interest recently in classical statistical physics.
While it was originally conjectured for continuous time, discrete state Markov processes 
in steady-state \cite{Barato:2015:UncRel},
it was later proved based on the large deviation technique \cite{Gingrich:2016:TUP,Horowitz:2017:TUR}.
An incomplete list of studies on  the TUR includes its generalizations to
finite-time statistics \cite{Dechant:2018:TUR,Pietzonka:2017:FiniteTUR,Horowitz:2017:TUR,Pigolotti:TURF},
Langevin dynamics \cite{Dechant:2018:TUR,Gingrich:2017,Hasegawa1,TUR-gupta,Hyeon:2017:TUR},
periodic dynamics \cite{Koyuk:2018:PeriodicTUR,Gabri}, broken time reversal symmetry systems 
\cite{Udo:TURB,Saito,Hyst}, as well as derivations of
trade-off relations for heat engines \cite{heat}.
Generalized versions of the TUR, which are based on the fundamental fluctuation relation were 
recently derived in Refs. \cite{VanTUR, Landi-PRL}. Furthermore, a TUR bound for quantum systems in 
non-equilibrium steady-state was 
obtained in Ref. \cite{LandiQ} using quantum information theoretic concepts.

For a two-terminal single-affinity system, the TUR connects the 
steady-state current
$\langle j\rangle$, its variance $\langle \langle j^2\rangle\rangle=\langle j^2\rangle-\langle j\rangle^2$, 
and the average entropy production rate $\langle \sigma\rangle $ in a nonequilibrium process \cite{Barato:2015:UncRel},
\bea
\frac{\langle \langle j^2\rangle \rangle}{  \langle j\rangle^2} \frac{\langle \sigma\rangle }{ k_B} \geq 2.
\label{eq:TUR0}
\eea
Here, $k_B$ is the Boltzmann constant.
This relation, which was originally derived 
based on Markovian dynamics, reduces to an equality in linear response. 
Away from equilibrium, Eq. (\ref{eq:TUR0}) describes the
trade-off between precision and dissipation:
A precise process with little noise is realized with high thermodynamic (entropic) cost.
Systems that obey this inequality \textit{satisfy the TUR}. 
TUR violations correspond to situations in which the left hand side of Eq.
(\ref{eq:TUR0}) is smaller than 2. We refer to this special bound as the TUR---contrasting it to generalized TUR bounds---and we
highlight that it is not universal and that it may be violated for certain processes \cite{BijayTUR}.  

Violations of the bound (\ref{eq:TUR0}) were theoretically 
predicted in Refs. \cite{BijayTUR,BijayH,Junjie} for charge and energy transport problems 
in single and double quantum dot junctions in certain parameter regimes, when the transmission function was structured in the bias window.
The first experimental interrogation of the bound (\ref{eq:TUR0}) was recently reported in Ref. \cite{TURNMR}
by probing energy exchange between qubits---albeit in the {\it transient} regime.
It was demonstrated in Ref. \cite{TURNMR} that this bound could be violated by tuning the energy exchange parameters (qubit-qubit coupling),
in line  with theoretical predictions, and while
satisfying the looser, generalized TUR bounds \cite{VanTUR, Landi-PRL, LandiQ}.

In what follows, we focus on the bound (\ref{eq:TUR0}), rather than on its generalized forms \cite{VanTUR, Landi-PRL}
or the looser quantum bound \cite{LandiQ},
since it is expected to be valid for quantum transport junctions with a constant
transmission probability \cite{BijayTUR}.
Considering (single-affinity) steady-state charge transport under an applied bias voltage $V$, 
dissipation is given by Joule's heating, 
$\langle \sigma \rangle = \langle j\rangle \frac{V}{T}$, with $T$ the temperature of the electronic system.
The inequality (\ref{eq:TUR0}) then simplifies to
\bea
 \beta V \frac{\langle \langle j^2\rangle \rangle}{  \langle j\rangle} \geq 2,
\label{eq:TUR1}
\eea
with $\beta= (k_BT)^{-1}$.
For convenience, we introduce the combination
${\mathcal Q} \equiv  \beta V \frac{\langle \langle j^2\rangle \rangle}{  \langle j\rangle}$,
which is a function of voltage and temperature. We refer to ${\mathcal Q}$ as the \textit{TUR ratio}. 


The TUR allows understanding of the trade-off between current fluctuations and entropy production.
Furthermore, verifying or violating Eq. (\ref{eq:TUR1}) provides insight into 
the underlying charge transport statistics as was discussed in Ref. \cite{BijayTUR}.
Atomic-scale junctions offer a rich playground for studying steady-state quantum transport at the nanoscale \cite{BookJS}. 
It was pointed out in Ref. \cite{BijayTUR} that in junctions with a constant transmission probability, Eq. (\ref{eq:TUR1})
should be valid. However, an experimental verification for this prediction is missing.
Furthermore, beyond the fundamental interest in thermodynamical bounds, it might be useful to examine the behavior of the TUR ratio.
We therefore ask here the following question:
Does the measure $\mathcal Q$ 
reveal useful, additional information about the transport process 
beyond what is separately contained in the current and its fluctuations?

The objective of this work is to study 
the TUR in charge-conducting atomic-scale junctions and
use this compound measure to learn about charge transport mechanisms in real systems.
Different realizations of gold atomic scale junctions depict
distinct differential conductance traces \cite{BookJS}.
Furthermore, corresponding shot noise measurements 
display pronounced anomalous characteristics at high voltage \cite{Natelson1, Natelson2, Ruitenbeek,Anqi}.
Here, we confirm the validity of the TUR [Eq. (\ref{eq:TUR1})] in steady-state 
using experimental data, in accord with theoretical  predictions \cite{BijayTUR}.
Furthermore, we argue that the TUR ratio can 
distill underlying transport mechanisms in atomic-scale junctions, which
may be convoluted at the level of the current and its noise. 
The ratio ${\mathcal Q}$ begins at the equilibrium value of 2. We show that its linear behavior in voltage
indicates the shared underlying quantum coherent dynamics.
In contrast, a quadratic term in voltage distinguishes nonlinear contributions beyond the constant transmission limit.

Altogether, this study (i) validates and verifies 
the TUR in atomic-scale junctions and (ii) illustrates
that the TUR can assist in diagnosing transport regimes.
Yet more broadly, this study bridges a gap between quantum transport junctions \cite{BookJS} and stochastic
thermodynamics \cite{Udoreview,England}, illustrating that thermodynamical bounds can be tested in
nanoscale systems, in the quantum domain, down to the level of atomic-scale electronic conductors.


\section{Theory}
\label{sec-mainR}

\subsection{TUR for normal shot noise}

We consider nanoscale conductors in the quantum coherent limit with 
a constant transmission function (ohmic conductors), $\tau=\sum_i\tau_i$, collecting the contribution of independent transmission 
channels. The electrical current and its noise, under the chemical potential $\Delta \mu = eV$, 
are given by \cite{Buttiker, Nazarov,BookJS}
\bea
\langle j\rangle &=& G_0V\sum_i\tau_i,
\nonumber\\
\langle \langle j^2\rangle \rangle &=& 
2k_BT G_0 \sum_i \tau_i^2 
\nonumber\\
&+& 
G_0 \sum_i \tau_i(1-\tau_i) \Delta \mu \coth \left(\frac{\Delta \mu}{2k_BT}\right).
\label{eq:currentnoise0}
\eea
We identify the electrical conductance, $G=G_0 \sum_i \tau_i$, and
the constant Fano factor $F=\sum_i \tau_i(1-\tau_i)/\sum_i \tau_i$. 
$G_0=2e^2/h$ is the quantum of conductance.
Note that the zero frequency spectral density of the noise, commonly denoted by $S(\omega=0)$, is 
defined a factor of 2 greater than the second cumulant of the noise.
The expression for the current noise combines the (zero voltage) thermal noise and the (zero temperature) 
shot noise. From these expressions we prepare the TUR ratio,
\bea
{\mathcal Q}= 2 + \frac{\sum_i \tau_i(1-\tau_i)}{\sum_i \tau_i}
\left[ \frac{\Delta \mu}{k_BT} \coth \frac{\Delta \mu}{2k_BT} -2 \right].
\label{eq:Qtau0}
\eea
Since $x \coth x\geq 1$, the bound (\ref{eq:TUR1}) is satisfied 
in the constant transmission limit, independent of voltage and temperature.
In the high bias limit, $\Delta \mu\gg k_BT$, this relation reduces to
\bea
{\mathcal Q}=2+F\beta |\Delta \mu|.
\label{eq:Qconst}
\eea
Note that in this limit the current noise is  $\langle \langle j^2\rangle \rangle=  
e |\langle j \rangle| F$, which is the quantum shot noise with the suppression factor $F$.
In contrast, in the limit of very low voltage we expand Eq. (\ref{eq:currentnoise0}) and get
\bea
{\mathcal Q}=
2 +
\frac{(\beta \Delta \mu)^2}{6} F 
+ {\mathcal O}((\Delta \mu)^4) 
+ \cdots.
\label{eq:lowV}
\eea
To study the latter expansion, one would need to inspect the current and its noise close to equilibrium.
For T=7 K, $\beta\approx 1500 $ (eV)$^{-1}$ and $|\beta V| <1$ requires scanning the noise for fine bias voltage
$V< 1$ mV. 
In the experiments reported below the voltage was
 scanned between 10 and 1000 mV, focusing on the examination of Eq. (\ref{eq:Qconst}).

Comparing Eq. (\ref{eq:currentnoise0}) to Eq. (\ref{eq:Qtau0}), we note that these expressions are closely related. 
However, we argue that the TUR ratio, Eq. (\ref{eq:Qtau0}), provides a beneficial representation of the scaled noise 
since (i) its development from the equilibrium value of 2 to the high voltage regime can be clearly observed, 
and
(ii) it brings the data together onto a universal curve.
In Sec. \ref{Sexp} we demonstrate these points on measured data.

\subsection{TUR for anomalous shot noise}

Measurements of shot noise in Au atomic-scale contacts reveal anomalous (nonlinear) characteristics at high voltage
\cite{Natelson1, Natelson2, Ruitenbeek,Anqi}.
These observations were interpreted in Ref. \cite{Anqi} based on a coherent quantum transport model
with two elements: The transmission function for electrons was assumed to be 
energy dependent, and the voltage drop on the electrodes was allowed to be asymmetric. 

For the Au atomic-scale junctions analyzed in this work,  $G\approx 1G_0$, 
and it is therefore sufficient to consider two channels \cite{BookJS,Anqi}:
a primary channel, which is almost fully open, and a secondary channel with a low transmission probability.
We  model the transmission function of the  dominant channel by the low order (linear) 
Taylor expansion
$\tau_1(\epsilon)= \tau_1 + \tau_1'(\mu)(\epsilon-\mu)$, where
$\tau_{1}$ is the constant value of the transmission function at the Fermi energy and
$\tau_1'(\mu)$ is the derivative of this function, evaluated at the Fermi energy.
The contribution of the secondary channel is minor, and
for its transmission function we use the constant approximation, 
$\tau_2(\epsilon)\sim \tau_2 \ll  \tau_1$. 
The partition of the bias voltage is quantified by the parameter $\alpha$,  with
$\mu_{L}=\mu + \alpha \Delta \mu$ and $\mu_R=\mu-(1-\alpha)\Delta \mu$;  $0\leq\alpha\leq 1$;
the bias voltage is symmetrically divided at the electrodes when $\alpha=1/2$.
It should be highlighted that the linear approximation for the transmission function
describes only a certain class of measurements, while other atomic-scale junctions display more complicated trends \cite{Anqi}.

The determinant for the energy (and  possibly voltage) dependent transmission function
in Au atomic-scale junctions could be quantum interference of electron waves with randomly-placed 
defects in the metal contacts \cite{Ruitenbeek}; our analysis does not presume the root for the functional form of $\tau(\epsilon)$.
The origin of the bias voltage asymmetry could be structural differences in the contact region at the left and right sides.
The mean-field parameter $\alpha$ emerges due
to underlying many-body effects, that is, the response of electrons in the junction to the applied electric field.

Based on these ingredients, expressions for the current (divided here already by voltage) 
and its noise were derived in Ref. \cite{Anqi},
\bea
\frac{\langle j \rangle}{V} &=& G_0 \sum_i\tau_i + G_0\tau_{1}'(\mu)(\alpha-1/2) \Delta \mu,
\nonumber\\
\langle \langle j^2 \rangle\rangle &\approx & 2
k_BT G_0 \sum_i \tau_i 
\nonumber\\
&+& 
G_0k_BT \sum_i \tau_i(1-\tau_i) \left[ \frac{\Delta \mu}{k_BT} \coth \left(\frac{\Delta \mu}{2k_BT}\right)
-2 \right]    
\nonumber\\
&+& G_0 (1-2\tau_1) \tau_1'(\mu) (\alpha-1/2) (\Delta \mu)^2.
\label{eq:Anqi}
\eea
The complete expression for the noise is included in the Appendix;
here we already took the low temperature limit $eV\gg k_BT$ and assumed that $|\tau'\Delta \mu|<1$ and 
 $\alpha \neq 1/2$.
For simplicity, we define a combined nonlinear
coefficient $\gamma\equiv (\alpha-1/2)\tau_1'(\mu)$, which has the physical dimension of inverse energy.
This coefficient conjoins the two elements that are responsible for
anomalous behavior: Many body electron-electron effects
(phenomenologically captured by $\alpha$) and an energy-dependent transmission probability.

We now write down the TUR ratio in the limit $eV \gg k_BT$ to the lowest order  in $|\gamma\Delta \mu| <1$,
for  the positive voltage branch, 
\bea
{\mathcal Q} \approx  2+  \Delta \mu  \beta F -\frac{2  \gamma\Delta \mu }{\sum_{i}\tau_i}
- \frac{\gamma \beta (\Delta \mu)^2}{\sum_i \tau_i} \left[ F + (2\tau_1-1) \right  ].
\nonumber\\
\label{eq:Qtau}
\eea
We can simplify this expression by noting that in our experiments (see Sec. \ref{Sexp}) $\beta F \gg |\gamma|$ 
and that $\tau_{2}\ll \tau_1$. We get
\bea 
\frac{F+ (2\tau_1 -1)}{\sum_i \tau_i}= 1 - \frac{2\tau_2^2}{\left(\tau_1 + \tau_2\right)^2} \approx 1,
\eea
which simplifies Eq. (\ref{eq:Qtau}) to
\bea
{\mathcal Q} \approx  2+  \Delta \mu \beta F  - \gamma \beta (\Delta \mu)^2.
\label{eq:QG}
\eea
This result is remarkable since the second, nonlinear term in voltage distills the nonlinear 
contribution $\gamma$. 
Since $\gamma$ could be positive or negative, the TUR ratio may show either a suppression or an enhancement from the
linear normal shot noise term. 
Further, $\gamma \Delta \mu $ could be comparable to the constant Fano factor $F$,
therefore the contribution of the quadratic term could be substantial.
In fact, the TUR could be violated at high voltage once $F<\gamma \Delta \mu$.

Altogether, we argue that presenting the noise as a TUR ratio is beneficial for elucidating transport processes.
According to Equation (\ref{eq:QG}): (i) the constant term, 2, 
represents the equilibrium value.
(ii) The linear term in voltage describes quantum suppressed-Poissonian dynamics and it
identifies the corresponding Fano factor. This term emerges from a quantum coherent transport process
with a constant transmission coefficient. 
(iii) The nonlinear term includes deviations from
the constant transmission limit and it reflects the departure from the simple quantum coherent picture,
with the involvement of many body effects ($\gamma$).

Equation (\ref{eq:QG}) is valid at high voltage, or correspondingly, low temperature, $\Delta \mu \gg k_BT$. 
In the Appendix we discuss the behavior of the TUR ratio for anomalous shot noise at high temperature.

\section{Analysis of atomic-scale gold junctions}
\label{Sexp}

\begin{figure}[t]
\hspace{-2mm}\includegraphics[scale=0.3]{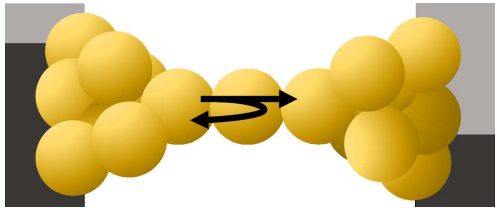}
\caption{An illustration of a gold atomic-scale junction realized with the break junction technique.
At high bias voltage and zero temperature, electron current is unidirectional, 
with the transmitted and reflected components illustrated by arrows. The dark (light) regions at
the left and right sides represent electron occupation (empty states).
The atomic configuration at the junction varies, and the junction is not necessarily spatially symmetric.
As a result, the applied voltage may be partitioned unevenly across the atomic-scale junction, quantified by the parameter $\alpha$.
}
\label{scheme}
\end{figure}

\begin{figure*}[t]
\hspace{-2mm}
\advance\leftskip-1.5cm
\includegraphics[scale=0.43]{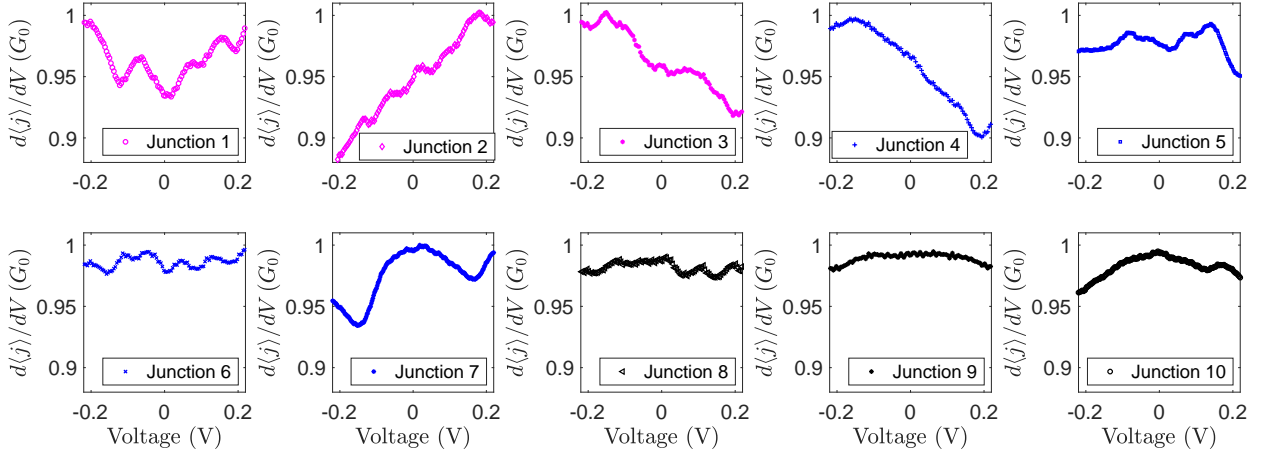}
\caption{Electrical conductance of 10 representative Au atomic-scale junctions formed using the
mechanically controllable break-junction technique.}
\label{conductance-exp}
\end{figure*}

\begin{figure*}
\hspace{-6mm}\includegraphics[scale=0.48]{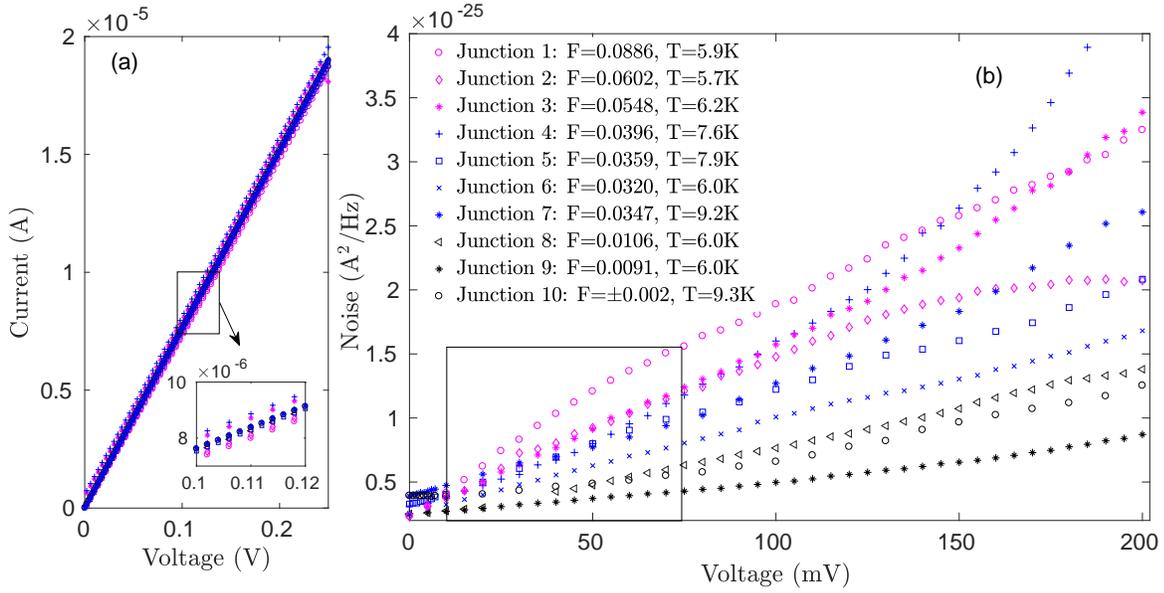} 
\caption{(a) Current as a function of voltage and (b) its noise
for each of the ten atomic-scale junctions of Fig. \ref{conductance-exp}.
The temperature (extracted from the equilibrium noise) and the
constant Fano factor (obtained from the low-voltage noise)
are presented in the legend.
The noise is approximately linear in bias voltage for most junctions between 10 mV$ <V< $75 mV, indicated by the region enclosed in a rectangle. 
For junction 10, the current noise does not display a normal shot noise region,
resulting in an undetermined (very small) Fano factor. 
}
\label{raw}
\end{figure*}

\begin{figure*}[t]
\hspace{-10mm}\includegraphics[scale=0.43]{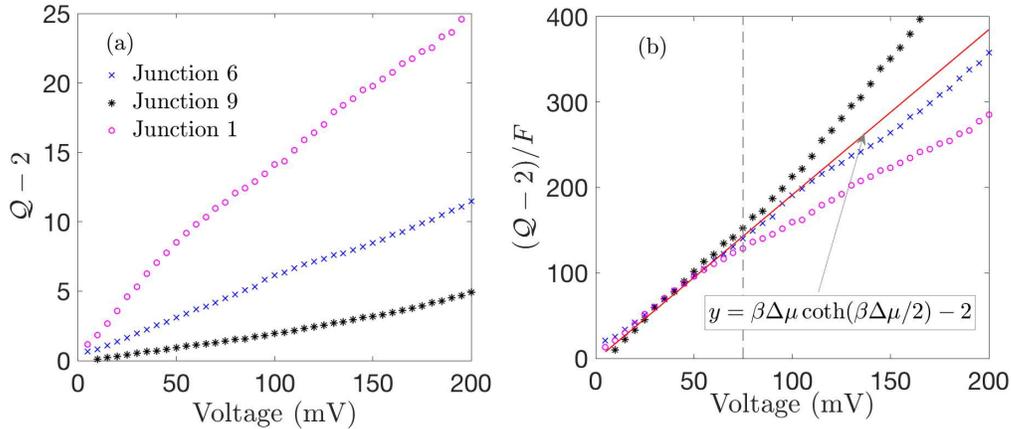} 
\caption{(a) TUR ratio as a function of voltage for junctions with approximately a constant transmission function 
(junctions 6 and 9), as well as for junction 1, which shows an energy dependent transmission function. 
(b) 
By plotting the data as $({\mathcal Q}-2)/F$,
the different curves collapse onto the universal function (\ref{eq:Qtau0}) before the dashed line, 
which marks the bias voltage 75 mV, beyond which deviations from the universal function show. 
The temperature, calculated from the equilibrium noise, is $T=6.0 $ K.
$G=G_0\sum\tau_i$ was obtained from the zero-voltage differential conductance.
The constant Fano factor was deduced from the normal shot noise regime, based on the behavior of the differential conductance and the
noise.   
Not all voltage points in the noise trace were measured in the electrical current.
However, since the current is highly linear in voltage, 
we performed a linear  interpolation for the current-voltage curve and added the few missing points in between; 
we only interpolate for voltages higher than 10 mV.
}
\label{FQ1}
\end{figure*}

\begin{figure*}[t]
\hspace{-8mm}\includegraphics[scale=0.55]{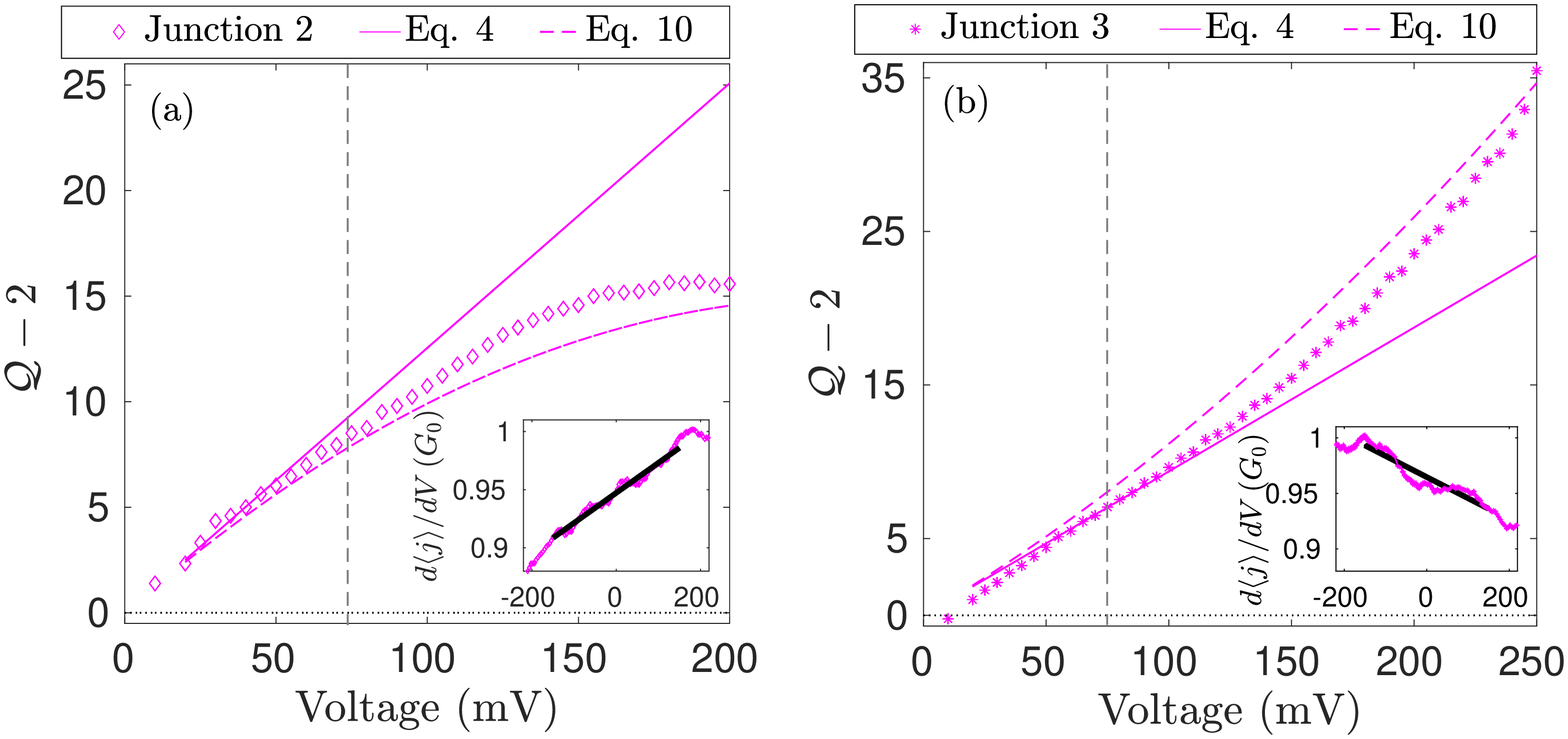} 
\caption{TUR ratio in junctions 2 and 3 for which the
differential conductance (inset), and therefore the transmission function can be approximated by a linear function.
Measurements are compared to (full line) the constant transmission TUR, Eq. (\ref{eq:Qtau0}) and to the
(dashed line) nonlinear expression, Eq. (\ref{eq:QG}). Vertical lines at 75 mV mark the onset of the anomalous regime.
The insets display the differential conductances for the two junctions, with a linear fit (dark full line) performed around
zero voltage yielding the nonlinear coefficients
(a) $\gamma=0.129$ 1/eV and (b) $\gamma=-0.095$ 1/eV.
}
\label{FQ2}
\end{figure*}

We test the theoretical expressions for the TUR ratio with measurements 
of the current and current fluctuations in Au atomic-scale junctions. The atomic junction and scattering processes are  illustrated in
Fig. \ref{scheme}.
We use the mechanically-controllable break junction technique at cryogenic conditions to form an ensemble of
atomic-scale junctions \cite{Ofir}. By repeatedly breaking and reforming the junction, realizations 
with somewhat different atomic configurations are generated, supporting a range of conductance values of $G\approx 1G_0$. 
Differential conductance measurements are performed for each junction, as well as current-voltage
traces and shot noise measurements. 
For gold atomic-scale junctions, proximity-induced 
superconductivity measurements \cite{GoldS} and shot-noise analysis \cite{oren13,Vardimon16}
suggest that a single channel dominates the conduction with a nearly perfect coupling to the metals \cite{BookJS}.

The data that we analyze: the differential conductance, current-voltage characteristics and the shot noise
for different atomic-scale junctions
is shown in Figures \ref{conductance-exp} and \ref{raw}.
Based on  Fig. \ref{raw}, we classify three regimes in the current noise:
(i) The close-to-equilibrium, or the low voltage regime. 
In this case, $eV$ is lesser or equal to the thermal energy and the thermal noise is prominent; if $T<10$ K, $V<$ 5 mV.
The TUR in this regime [see Eq. (\ref{eq:lowV}) and the Appendix] is not probed in our work.
(ii) Normal shot noise regime, 10 mV$<V\lesssim 75$ mV.
In this region the current noise follows the standard-normal shot noise expression, and it is linear in voltage.
(iii) Anomalous shot noise regime, around $V>$ 75 mV. 
In this region the shot noise displays anomalous trends as it is no longer linear in voltage.

We extract the zero-voltage electrical conductance, $G=G_0\sum_i \tau_i$, 
from the differential conductance at the zero voltage.
The temperature is verified from the equilibrium noise $\langle \langle j^2\rangle\rangle=2k_BTG$, 
and is in the range of $T=5-10$ K. 
The constant Fano factor $F$ is obtained by fitting the shot noise at low voltage (typically, $V< 20$ mV) 
to Eq. (\ref{eq:currentnoise0}) and dividing by the zero-voltage electrical conductance.

In Fig. \ref{conductance-exp}, we display the differential conductance of 10 representative junctions. 
While in some cases the differential conductance is about constant with voltage, other junctions
explicate a more significant variability of $d\langle j \rangle/dV$ with voltage, 
indicating deviations from Eq. (\ref{eq:currentnoise0}). 
We use different colors for different junctions roughly grouped 
according to their constant Fano factor. 
In Fig. \ref{raw}(a), we display the currents for these 10 junctions, 
which are nearly ohmic throughout.
We further present the current noise in Fig. \ref{raw}(b), which largely deviates from a linear behavior beyond $V \sim 75$ mV
\cite{Anqi}. Recall that $\langle \langle j^2\rangle \rangle \propto |V|$ for ohmic conductors. Deviations from
this trend are referred to as the ``anomalous shot noise". This effect was the focus of Ref. \cite{Anqi}.


To test Equation (\ref{eq:Qtau0}), we select junctions 6 and 9 that display approximately 
a constant differential conductance (thus a constant transmission function),
see Fig. \ref{conductance-exp}
(we exclude junction 8 since its noise measurements were missing values in the normal shot noise voltage regime). 
To contrast it, we also analyze junction 1 for which the differential conductance 
varies more substantially with voltage. 
While the temperature was similar for the three junctions  ($T=6.0\pm$ 0.1 K),
the Fano factor was quite distinct, varying between 0.1 to 0.01.

We present the TUR ratio (after subtracting the equilibrium value), ${\mathcal Q}-2$ 
of the three junctions in Fig. \ref{FQ1}(a).
As expected, ${\mathcal Q}\geq2$ throughout, 
Furthermore, by plotting the ratio $({\mathcal Q}-2)/F$ in Fig. \ref{FQ1}(b) we demonstrate that the
measurements collapse on the universal function $\beta V \coth (\beta V/2)-2$ 
up to around 75 mV for junctions 6 and 9. 
In fact, when $T\sim 6$ K ($\beta \sim 2000$ 1/(eV)), 
$|\coth \beta \Delta \mu/2|\approx 1$ beyond 5 mV, thus
arriving at Eq. (\ref{eq:Qconst}) with
$({\mathcal Q}-2)/F \approx \beta |\Delta \mu|$.
The fact that the data agrees with Eq. (\ref{eq:Qtau0}) is not surprising, since
$F$ was extracted from the noise formula (\ref{eq:currentnoise0}). However, it is advantageous to present the data in this manner:
The combination $({\mathcal Q}-2)/F$ illustrates the common quantum coherent transport mechanism underlying the 
shot noise in these junctions for $V\lesssim 75$ mV. 

We test Eq. (\ref{eq:QG}), which describes deviations from the universal form
by studying junctions 2 and 3. 
In these two cases, the differential conductance  
approximately follows a linear line---corresponding to the theoretical model behind Eq. (\ref{eq:Anqi}).
Junction 4, which was analyzed in Ref. \cite{Anqi} suffers from more significant $1/f$ noise contribution, 
and we therefore do not include it. 

The slope of the differential conductance provides the coefficient $2\gamma$, 
which is adopted in Eq. (\ref{eq:QG}) to calculate the TUR ratio.
Results are displayed in Fig. \ref{FQ2}. 
The TUR ratio agrees with the constant transmission expression
(\ref{eq:Qtau0}) up to $\approx$ 75 mV. However, as we increase the voltage, the 
nonlinear expression (\ref{eq:QG}) provides a better description of the curved TUR
function with the quadratic coefficient $\beta \gamma$. 
We retrieve $\gamma \sim 0.13$ (eV)$^{-1}$ for junction 2. 
Given that $F=0.06$ for this junction, the TUR
could be violated at high voltage, $V>0.6$ V. 
However, at this bias voltage, one would need to consider higher order 
$\gamma \Delta \mu$ terms in the expansion (\ref{eq:QG}).
For junction 3, we obtain $\gamma \sim -0.095$ (eV)$^{-1}$, indicating the enhancement of the shot noise
relative to the normal shot noise regime. 
The lowest-voltage TUR ratio for junction 3 seemingly violates the TUR.
However, this point suffers from large relative error (since both the current and the noise are small), and
we cannot draw conclusions based on this single observation.
Careful measurements of the current and its noise at low voltage, $V<10$ mV,
would allow the analysis of the TUR close to equilibrium, as discussed in the Appendix. 



\section{Conclusions}
\label{sec-summ}

Cost-precision entropy-fluctuation trade-off relationships are
fundamental to understanding nonequilibrium processes.
In this work, we focused on steady-state charge transport in atomic-scale junctions, a process which is essentially quantum coherent. 
Based on experimental data and theoretical derivations, 
we show that the TUR (\ref{eq:TUR0}) is satisfied in this system, even when the noninteracting electron
picture is corrected to include many-body effects (in a mean-field form). 
The generalized quantum TUR \cite{LandiQ} is factor of 2 looser that this bound, and is obviously satisfied in our system.
Our work illustrates that the TUR bound is advantageous for exploring the fundamentals of 
transport processes: The TUR ratio is developed from the equilibrium value, and it therefore 
identifies far-from-equilibrium effects.
Indeed,
organizing the current noise as a TUR ratio is beneficial to understanding the charge transport problem. This is 
clearly observed by the evolution of the noise from the linear-universal behavior at intermediate voltage
to the anomalous regime at high voltage.

More broadly, our study illustrates that atomic-scale junctions offer a rich test bed for studying
theoretical results in stochastic thermodynamics--- while extending these predictions to the quantum domain.
As such, our combined theory-experiment analysis presents a step into consolidating the quantum transport and
statistical thermodynamics research endeavors.

The TUR for charge transport in steady-state, 
Eq.  (\ref{eq:TUR1}), can be violated once the transmission function is structured
with sharp resonances of width smaller than the thermal energy \cite{BijayTUR}.
This situation might be realized in molecular junctions at room temperature and at low voltage.
Specifically, quantum dot structures offer a rich playground
for studying the suppression of electronic noise in nanodevices. 
Experiments that directly probe the behavior of high order moments of the current \cite{Ensslin1,Ensslin2,Haug} 
could be used to examine thermodynamical bounds \cite{BijayTUR,Junjie}.
Future work will be focused on the behavior of the current noise and the associated TUR in 
many body systems such as transport junctions with pronounced electron-vibration coupling.

\begin{acknowledgments}
DS acknowledges the Natural Sciences and Engineering Research Council (NSERC) of Canada Discovery Grant 
and the Canada Research Chairs Program. 
The work of HMF was supported by the NSERC Postgraduate Scholarships-Doctoral program. 
OT appreciates the support of the Harold Perlman family, and acknowledges funding by a research grant from 
Dana and Yossie Hollander, the Israel Science Foundation (grant number 1089/15), and the Minerva Foundation 
(grant number 120865). 
\end{acknowledgments}

\vspace{4mm}
\renewcommand{\theequation}{A\arabic{equation}}
\setcounter{equation}{0}  

\section*{Appendix: Expansion of the TUR ratio close to equilibrium}

Equations  (\ref{eq:Qconst}) and (\ref{eq:QG}), which we used to explain exponential data,
were derived in the limit of high bias voltage, $eV\gg k_BT$. 
Here we study the complementary limit of high temperature (or low voltage) 
and discuss the possible violation of the TUR in this regime. 

The electric current and its noise can be formally expanded in orders of the applied bias voltage as
\bea
\langle j \rangle &=& G_1 V  + \frac{1}{2!} G_2 V^2 + \frac{1}{3!} G_3 V^3 +...
\nonumber\\
\langle \langle j^2\rangle \rangle  &=& S_0 + S_1 V + \frac{1}{2!} S_2 V^2 +    \frac{1}{3!} S_3 V^3 +... \,
\label{eq:expand}
\eea
Here, $G_1$ is the linear conductance,  $G_2$, $G_3$,...  are the nonlinear coefficients in the 
current-voltage expansion.
Similarly, $S_0$ is the equilibrium (Johnson Nyquist) noise, and $S_1$, $S_2$, are the
nonequilibrium noise terms.
We substitute these expansions into Eq. (\ref{eq:TUR1}) and get  \cite{BijayTUR},
\bea
\beta V \frac{\langle \langle j^2 \rangle \rangle}{\langle j \rangle} &=& \frac{\beta}{G_1} S_0 +
\frac{\beta V}{G_1} \Big[ S_1 - \frac{S_0 \, G_2}{2\, G_1}\Big] + \frac{\beta V^2}{G_1} \times \nonumber\\
&&\Big[ \frac{S_2}{2} - \frac{S_0 G_3}{6 G_1} + \frac{S_0 G_2^2}{4 G_1^2} - \frac{S_1 G_2}{2 G_1}\Big]
+ {\mathcal O}(V^3) + \cdots\nonumber \\
\label{eq:TURex}
\eea
We make use of the fluctuation-dissipation (Green-Kubo) relation, 
$S_0=2k_BTG_1$, and the first of the Saito-Utsumi relationships \cite{SaitoU},
$S_1=k_BTG_2$, both resulting from the fluctuation relation \cite{fluctRev},
and reduce  Eq. (\ref{eq:TURex}) to
\bea
\beta V \frac{\langle \langle j^2 \rangle \rangle}{\langle j \rangle} =
2 + \frac{V^2}{3S_0} \Big[ 3 S_2 - 2 k_B T G_3\Big]+ {\mathcal O}(V^3) + \cdots
\nonumber\\
\label{eq:TUR}
\eea
We now introduce the expansion of the TUR ratio around equilibrium,
\bea
{\mathcal Q} = 2 + Q_2 (\beta V)^2 +  Q_3 (\beta V)^3 + \cdots
\label{eq:Q}
\eea
Here, $Q_2$, $Q_3$,... are coefficients of the TUR ratio, and they depend on internal parameters and temperature.
Note that $Q_1$ is missing in this expansion \cite{BijayTUR}.
A negative $Q_2$ identifies TUR violation in the second order of voltage.

We now specify this analysis to the gold atomic-scale junctions with $G\sim 1 G_0$, as described in Sec. \ref{Sexp}
and in Ref. \cite{Anqi}.
To model quantum coherent transport in Au junctions we assume that: 
(i) The transmission function, which describes the probability for electrons 
to cross the junction, is linear in energy.
(ii) The bias voltage is divided asymmetrically at the contacts, quantified by the parameter $\alpha$.
(iii) Two channels contribute to the transmission, a dominant one which is almost fully open and 
a secondary channel with a small transmission coefficient $\tau_2\ll\tau_1 $.
(iv) We take into account the variation of the transmission function with energy for the dominant channel only.
Using this setup, the charge current is given by \cite{Anqi}
\bea
\langle j \rangle = \frac{2e}{h}  \sum_{i=1,2}\tau_{i} \Delta \mu
+ \frac{2e}{h}  \tau_1'(\mu) \left(\alpha-\frac{1}{2}\right) ( \Delta \mu)^2.
\label{eq:current}
\eea
The linear conductance and the nonlinear coefficients can be collected as
\bea
G_1&=&G_0\sum_i{\tau_{i}}, \,\,\,\, G_2= G_0 \tau_1'(\mu)(\alpha-1/2)(2e), 
\nonumber\\ 
G_3&=&0.
\eea
The shot noise is given by \cite{Anqi}
\begin{widetext}
\bea
\langle \langle j^2\rangle \rangle/ G_0&=&
2k_B{T}\sum_i\tau_{i}^2
+ \sum_i\tau_{i}(1-\tau_{i})\Delta \mu\coth\left(\frac{\Delta \mu}{2 k_B {T}}\right)
\nonumber \\
&+& 2k_B{T}\tau_{1} \tau_1'(\mu)\Delta \mu (2\alpha-1)
+  2k_B{T}[\tau_1'(\mu)]^2\frac{\pi^2k_B^2{T}^2}{3}+k_B{T}[\tau_1'(\mu)]^2\left[\alpha^2 (\Delta \mu)^2 +(1-\alpha)^2(\Delta \mu)^2\right] \nonumber \\
&+&\coth\left(\frac{\Delta \mu}{2 k_B {T}}\right)(1-2\tau_{1}) \tau_1'(\mu)\left(\alpha-\frac{1}{2}\right)(\Delta \mu)^2
\nonumber \\
&-&\coth\left(\frac{\Delta \mu}{2 k_B {T}}\right)[\tau_1'(\mu)]^2\left[\Delta \mu\frac{\pi^2k_B^2{T^2}}{3}+\frac{1}{12}(\Delta \mu)^3+
\left(\alpha-\frac{1}{2}\right)^2 (\Delta \mu)^3\right],
\label{eq:noise}
\eea
\end{widetext}
with the first three coefficients  (\ref{eq:expand}),
\bea
S_0&=& 2k_BT G_0\sum_{i}\tau_{i},\,\,\,\,\
S_1= k_BTG_0 (\alpha-1/2) \tau'_1(\mu)(2e),
\nonumber\\
S_2&=& G_0e^2\frac{\sum_i \tau_{i}(1-\tau_{i})}{3k_BT} - k_BTe^2 G_0\frac{(\pi^2-6)(\tau'_1)^2}{9}.
\label{temp-relation}
\eea
We can now verify 
the fluctuation-dissipation relation, $S_0=2k_BTG_1$, as well as the first of the Saito-Utsumi relations,
$S_1=k_BTG_2$. Indeed, though $\alpha$ (phenomenologically) builds on many-body effects, we
can still write down the Levitov-Lesovik formula for the cumulant generating function
and show that it satisfies the exchange steady-state fluctuation
symmetry \cite{fluctRev}.
Since $G_3=0$, the series for the TUR ratio, Eq. (\ref{eq:Q}), reduces to 
\bea
\beta V \frac{\langle \langle j^2 \rangle \rangle}{\langle j \rangle} =
2 + V^2\frac{S_2}{S_0} + {\mathcal O}(V^3) + \cdots,
\label{eq:TUR2}
\eea
%
Substituting $S_0$ and $S_2$ we get
\bea
\beta V \frac{\langle \langle j^2 \rangle \rangle}{\langle j \rangle} = 2 +
\frac{(eV)^2}{6}\left[\beta^2 F    - \frac{(\tau'_1)^2 (\pi^2-6)}{3\sum_{i}\tau_{i}}  \right]  
+ {\mathcal O}(V^3) 
\nonumber\\
\label{eq:TUR3}
\eea
This expansion  is valid only close to equilibrium, and as such it is complementary to  Eq.
(\ref{eq:Qtau}), which was derived at high voltage.

Based on Eq. (\ref{eq:TUR3}),
can we  observe violations of the TUR in atomic-scale junctions---in the low-voltage regime?
For Au atomic-scale junctions $\sum_i\tau_i\approx 1$, $F\sim 0.01-0.1$ and $\tau_1'<0.1$ 1/(eV).
Therefore, the TUR is satisfied even at high temperatures, $T=1000$ K.
However, in systems with a small transmission coefficient, $\tau\ll 1$
(possibly molecular junctions),
TUR violations could be expected at high temperature once $\tau < (\tau_1')^2(k_BT)^2$. 

Altogether, the TUR is satisfied in atomic-scale junctions given that the
transmission coefficient is constant (energy independent). 
Furthermore, as discussed in Ref. \cite{BijayTUR}, 
while the single resonance level model can only display very weak TUR violations at high temperature,
double-dot models could break the TUR quite substantially  
depending on the inter-site coupling and the metal-dots hybridization energy.



\end{document}